\documentclass[mypaper,7pt,twoside]{CoAst}
\usepackage{epsf,graphicx,fancyhdr}
\renewcommand{\chaptermark}[1]{\markboth{#1}{}}

\newcommand{\kopf}{\small\itshape Comm. in Asteroseismology \\ Contribution to the Proceedings of the 38$^{th}$\,LIAC\,/\,HELAS-ESTA\,/\,BAG, 2008
}

\newcommand{\Authors}[1]{\begin{center}\normalsize\bf\sf #1 \end{center}}

\renewcommand{\author}[1]{\begin{center}\normalsize\bf\sf #1 \end{center}}
\newcommand{\Address}[1]{\begin{center}\small\sf #1 \end{center}}

\DeclareGraphicsExtensions{.eps,.jpg}

\newcommand{\Session}[1]{{\vspace{3mm}\small \noindent  \hspace*{3mm} Session: } #1 \normalsize}

\newcommand{\Objects}[1]{{\vspace{0mm}\small \noindent  \hspace*{3mm} Individual Objects: } \small #1 \normalsize}

	\newcommand{\four}{\small Observed frequencies in pulsating massive stars}

\renewenvironment{abstract}{\section*{Abstract}\normalsize\sf}{}
\newcommand{\References}[1]{\begin{flushleft}{\large References\\}\vspace*{2mm}\small #1 \end{flushleft}}

\newcommand{\chapterCoAst}[2]{\chapter[\sf\normalsize #1\\ \footnotesize \hspace*{5mm}by #2 \sf\normalsize][]{#1\\}\rhead[\fancyplain{}{\sf\footnotesize \center{#1}}]{\fancyplain{}{\sffamily\thepage}}\lhead[\fancyplain{\kopf}{\sffamily\thepage}]{\fancyplain{\kopf}{\sf\footnotesize \center{#2}}}}

\newcommand{\figureCoAst}[5]{\begin{figure}[#4]
\centering
\includegraphics*[#5]{#1}
\caption{#2}
\label{#3}
\end{figure}}

\renewcommand{\chaptername}{}

\newcommand{\acknowledgments}[1]{\vspace*{5mm}\noindent  \textbf{Acknowledgments.} #1}

\def\rfr{\smallskip\par\noindent
        \hangindent=7truemm
        \hangafter=1}

\begin{document}
\sf

\chapterCoAst{Ground-based observations of O and B stars}
{K.\,Uytterhoeven} 
\Authors{K.\,Uytterhoeven$^{1,2}$} 
\Address{
$^1$ Instituto de Astrof\'{\i}sica de Canarias, Calle Via L\'actea s/n, E-38205 La Laguna (TF), Spain \\
$^2$ INAF-Osservatorio Astronomico di Brera, Via E. Bianchi 46, I-23807 Merate, Italy}

\noindent
\begin{abstract}
Ground-based observations are a strong tool for
asteroseismic studies and even  in the era of  asteroseismic space
missions they continue to play an important role.  I will report on
the activities of the CoRoT/SWG Ground-Based Observations Working
Group, discuss the observational efforts of the Open Cluster
campaigns and  the search for the origin of  extra line-broadening in massive OB
stars.
\end{abstract}

\Session{ \four } 
\Objects{HD\,180642, HD\,50209, HD\,181231, HD\,49330, NGC\,3293, NGC\,6910, NGC\,884, NGC\,1893, NGC\,869} 

\section*{The role of ground-based observations in asteroseismic studies}
Over the last decades large observational efforts have been undertaken
to carefully monitor the pulsational variability of massive B stars,
resulting in a breakthrough in seismic modeling.  Also, several
observational initiatives are taken to open up new horizons, such as
the study of B-type pulsators in clusters and pulsational studies of
more evolved massive OB stars. The important key ingredients for an
asteroseismic study are precise pulsational frequencies, accurate
identified pulsation modes, and strong contraints on physical
parameters. To obtain these, a large amount of observing time is
required as continuous time-series are needed to unravel
beat-frequencies. The best we can do from the ground are multi-site
campaigns. Data from space have the additonal advantage of a good time
sampling and phase coverage, as well as a higher precision. However,
the ground-based observations continue to play an important role as
simultaneous ground-based data are complementary to the 'white light'
space data.  Multi-colour photometry provides information on amplitude
ratios and phase shifts, which are used to identify the degree $l$,
while high-resolution spectroscopy allows the detection of high-degree
modes and the identification of both $l$ and $m$ values.

\section*{The ground-based CoRoT support observations}
\begin{table}
\caption{Observatories, instruments and telescopes involved in the CoRoT preparatory and simultaneous ground-based observations.}
\label{instruments}
\centering
\begin{tabular}{ll|ll} \hline  \hline
\multicolumn{2}{c|}{preparatory} & \multicolumn{2}{c}{simultaneous} \\ \hline
Obs. & Instrument(s) & Obs.& Instrument \\ \hline
ESO & FEROS@1.52m&        CAHA &  FOCES@2.2-m \\	              		
KO & $V$@0.5m&         	  ESO  & FEROS@2.2m \\	              	
OHP & Elodie@1.93m   &	       &  HARPS@3.6m \\  	    
    & Aurelie@1.52m  & 	  MJUO & HERCULES@McLellen \\
OPM & NARVAL@TBL     &      OHP & Sophie@1.93m \\  
ORM & P7@Mercator    &	    OPM & NARVAL@TBL \\	  	    
    & SARG@TNG       &	    SNO & $uvby$@0.9m \\      		
OT & CCD@STARE       &	    SPMO & $uvby$@1.5m \\    	    		
SAAO & $UBV$@(0.5m,0.75m)&      & \\
SLN & FRESCO@0.91m &  &  \\
SNO & $uvby$@0.9m &   & \\
SPMO & $uvby$@1.5m &  & \\ \hline
\end{tabular}
\end{table}

\begin{table}
\caption{Overview of the amount of spectra  obtained from December 2006 until August 2008 in the framework of the simultaneous ground-based observational campaign of CoRoT targets with the FEROS, SOPHIE, FOCES and HERCULES spectrographs . The last column indicates the CoRoT run.}
\label{CoRoTtargets}
\centering
\begin{tabular}{ccccccc} \hline \hline
type & target &  FEROS & SOPHIE & FOCES & HERCULES & \\ \hline
$\delta$ Sct & HD\,50844  &  216 & & & & IR \\
 Be& HD\,50209 & 68 & & & & IR \\ 
 $\delta$ Sct & HD181555 & 343 & 66& 285 && LRc1\\
$\beta$ Cep & HD180642 & 213 & 35 & & &LRc1\\
Be& HD181231 & 72 & & & &LRc1\\
$\delta$ Sct & HD174966 & & & 134 &  &LRc1 \\
$\gamma$ Dor & HD\,49434 & 409 &711 & 75& 194 & LRa1\\ 
Be& HD\,49330 & 127 & & & & LRa1\\
 $\delta$ Sct & HD172189 & 176 & & & & LRc2\\
$\gamma$ Dor & HD171834 & 193 & 447 & 401 & 55 & LRc2 \\
Ap & HD171586 & 12 & & & & LRc2 \\ \hline
\end{tabular}  
\end{table}  

The CoRoT/SWG Ground-based Observations Working Group has played an
important role in preparing the CoRoT satellite mission.  Several mid-
and high-resolution spectrographs, multi-colour photometers, and a spectropolarimeter  at
different observatories\footnote{CAHA: Calar Alto Astronomical
Observatory (E); KO: Konkoly Observatory (HU); ESO: European Southern
Observatory, La Silla (CL); MJUO: Mount John University Observatory
(NZ); OHP: Observatoire de Haute Provence (F); OPM: Observatoire du
Pic du Midi (F); ORM: Observatorio Roque de los Muchachos (E); OT:
Observatorio del Teide (E); SAAO: South African Astronomical
Observatory (ZA); SLN: Osservatorio Serra La Nave (I); SNO: Sierra
Nevada Observatory (E); SPMO: San Pedro M\'{a}rtir Observatory (MX)}
were involved (see left panel of Table~\ref{instruments}) to
characterise and select suitable CoRoT targets (Poretti et al.\,2003,
2005).

Now the CoRoT satellite is in
successful operation, huge efforts are being made to guarantee the
simultaneous monitoring of a handful of selected $\beta$\,Cep,
$\delta$\,Sct, $\gamma$\,Dor and Be CoRoT targets from the ground.
Large Programme and normal observing proposals have been applied for,
and have been approved, with high-resolution spectrographs at ESO,
OHP, CAHA, and MJUO. Str\"omgren multi-colour information is provided
by telescopes at SPMO and SNO, and spectropolarimetry is done with NARVAL@TBL at OPM (right panel
Table~\ref{instruments}). Table~\ref{CoRoTtargets} gives an overview
of the targets and the amount of spectra that have been collected in
the framework of the simultaneous ground-based observations campaign
during CoRoT's first one-and-a-half years of operation\footnote{IR:
Initial Run (Jan.- Feb.\,2007); LRc1,LRc2: first (May-Oct.\,2007) and
second (Apr.- Sep.\,2008) Long Run in the center direction; LRa1: first
Long Run in the anti-center direction (Oct.\,2007 - Mar.\,2008)}.

\subsection*{First ground-based results of B-type CoRoT targets}
The $\beta$\,Cep star HD180642 (B1.5III, M~10$M_{\odot}$) was one of
the asteroseismic targets in CoRoT's LRc1.  Analysis of a dataset consisting of
507 high-quality multi-colour photometric data, obtained with the
Mercator telescope and with photometers at KO, SPMO, and SNO, and 280
high-resolution high S/N spectra (FEROS, SOPHIE, Aurelie) confirmed
the presence of a dominant radial mode, and revealed evidence for at
least two additional non-radial pulsations modes, including a possible
high-order g-mode (Uytterhoeven et al.\,2008; Briquet et al.\,2009, these proceedings).

No frequencies were found in the 68 FEROS spectra of the late-B type
Be star HD\,50209 (B8IV, v$\sin$i$=209$km s$^{-1}$), observed with
CoRoT in LRa1 (Guti\'errez-Soto \& Neiner, private comm.).  On
the other hand, photometric data, consisting of HIPPARCOS satellite
data, ASAS3 data and Str\"omgren $uvby$ data obtained at SNO, reveals
one frequency 1.47(2) $\pm$ 1d$^{-1}$ (Guti\'errez-Soto et al.\,2007).

No variable signals were detected in the HIPPARCOS data and SNO
Str\"omgren $uvby$ magnitudes of HD\,81231 (B5IV) (Guti\'errez-Soto et al.\,2007).  The Be star
was observed with CoRoT in LRc1. A line-profile analysis, based on 72
FEROS spectra, lead to the detection and identification of one $l=3$
mode (Guti\'errez-Soto \& Neiner, private comm.).

Guti\'errez-Soto et al.\, (2007) report the detection of frequencies in
the 2-3d$^{-1}$ domain in HIPPARCOS and SNO $uvby$ data for the
early-B type Be star HD\,49330. From a total of 127 FEROS and 41
NARVAL spectra it seems that this CoRoT target, observed in LRa1,
pulsates in several short-period variations (frequencies $>$
11d$^{-1}$), associated to high-degree modes ($5 < \ell < 7$), as well
(Floquet, Guti\'errez-Soto \& Neiner, private comm.). Longer-term
variations of the order of 50 days are also detected and can be
associated to the envelope of the Be system (Floquet, private comm.).

\subsection*{Future}
Analyses of the ground-based time-series in combination with the CoRoT
data are in progress, and the preliminary results look promising. To
continue the huge observational ground-based effort, there is a need
for (time on) high-resolution spectrographs. Even with the current
Large Programmes in progress, we need to keep convincing time
allocation committes the need for strings of consecutive nights and a
huge amount of observing time, which is not always straightforward
given the competition and the small amount of available
high-resolution spectrographs. Moreover, two of the instruments
intensively used for the ground-based support observations so far,
FOCES (R=40000) and FEROS (R=48000), will not be available anymore to
the community soon. This currently leaves us with the high-resolution
spectrographs HARPS (R=80000), SOPHIE (R=70000), HERCULES (R=35000)
and FIES (R= 46000 or 67000). Despite the two new spectrographs that will
be operational in the near future, HERMES (R=40000--90000) and SONG
(R=100000), it will be wise to look for alternatives to secure
 continuous ground-based high-resolution spectroscopic time-series in
the future.

\section*{The open cluster campaigns}
After successful multi-site campaigns on isolated $\beta$\,Cep stars,
a new challenge was found in performing asteroseismic studies of
$\beta$\,Cep stars in open clusters. A multi-site campaign using CCD
photometry was set up by Andrzej Pigulski et al.  The advantages of
such a study are clear: the cluster members are supposed to have the
same age and metallicity, which puts serious constraints on the
asteroseismic models. Moreover, as the CCD field contains several
cluster members, several pulsators can be studied simultaneously.  The
Open Cluster campaigns are being carried out in two phases:
observations in 2005--2007 were dedicated to NGC\,3293, NGC\,6910 and
NGC\,884 ($\chi$ Per), while NGC\,1893 and NGC\,869 (h Per) are being observed
in 2007-2009.

\subsection*{Preliminary results}
\figureCoAst{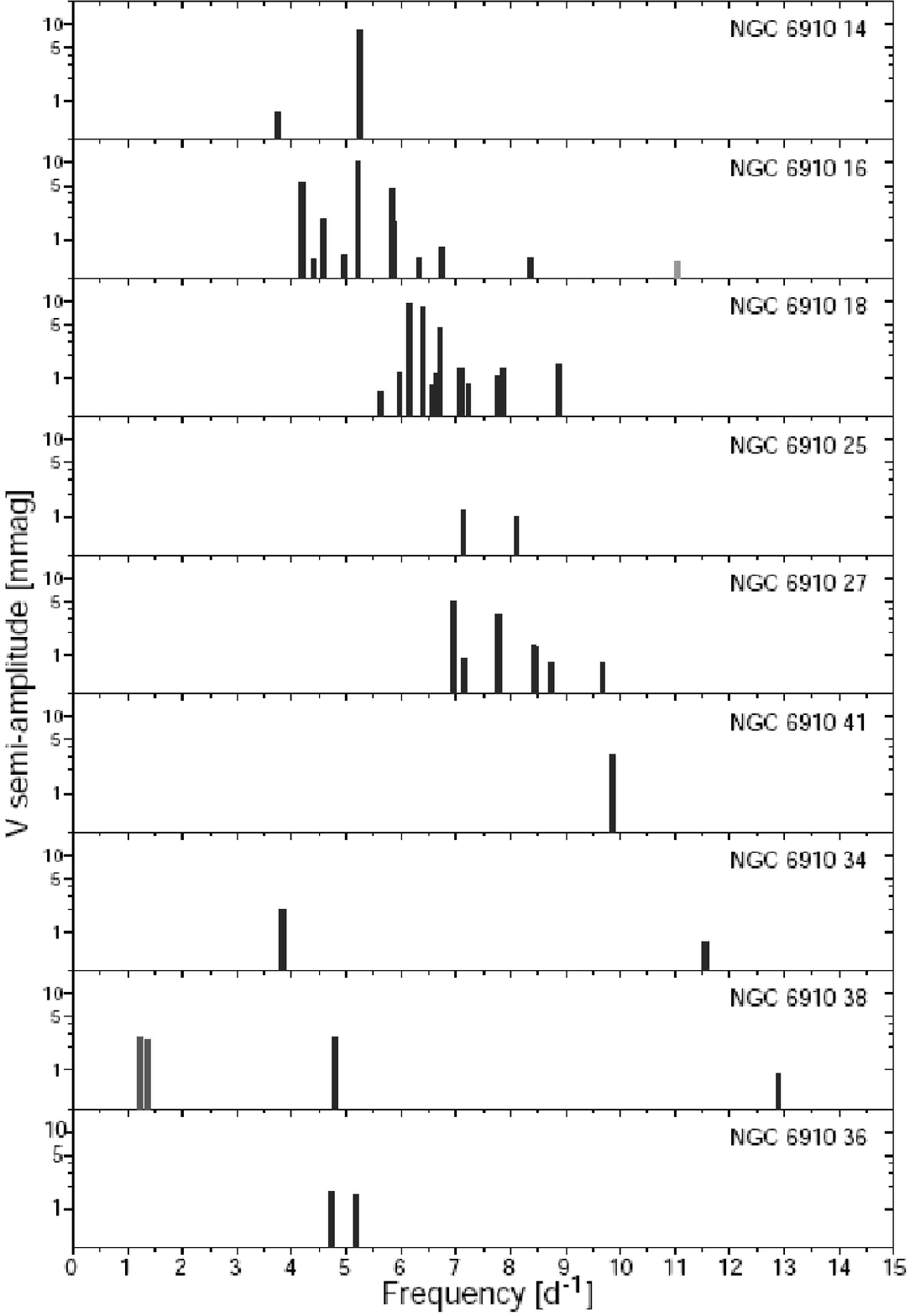}{Frequency spectra of the nine
detected $\beta$\,Cep variables in the open cluster NGC\,6910 (taken from Pigulski 2008).}{NGC6910}{b}{clip,angle=0,width=100mm}

The Southern cluster NGC\,3293 has been observed in January-May 2006
from seven sites. When combining all the
data, a detection level of 0.5mmag is expected. Handler et
al. (2007) reported on the first results obtained from the SAAO data
only: the cluster contains at least ten multiperiodic $\beta$\,Cep
stars and a few faint variable B stars with periods between 8-12
hours. The latter variables are very interesting from a theoretical
point of view, as their periods are shorter than typical Slowly Pulsating
B(SPB)-type pulsations and longer than the typical periods of
$\beta$\,Cep stars.

For the multi-site campaign on NGC\,6910 and $\chi$ Per, twelve sites in
ten countries on three continents were involved. The main telescopes were
the 0.6m telescope at Bia\l{}kow (PL), the Mercator telescope at ORM
(E) and the 0.8m telescope at the Observatory of Vienna University
(A).  We refer to Saesen et al. (2008; 2009, these
proceedings) for a detailed description of all the telescopes
involved, as well as for the preliminary results on the cluster $\chi$
Per. Analysis of the Bia\l{}kow data of NGC\,6910 resulted in the
confirmation of four known $\beta$\,Cep stars and the detection of five new
$\beta$\,Cep variables (Pigulski et al.\,2007; Pigulski 2008). For
some of them (e.g. NGC\,6910~18 and NGC\,6910~16) up to ten or more
modes are observed (see Fig.~\ref{NGC6910}).  Interestingly, both $p$-
and $g$-modes seem to be present in NGC\,6910~38 (Pigulski 2008).

Sites involved in the multi-site campaigns of the open clusters
NGC\,1893 and h\,Per, executed in 2007-2009, are Xinglong (CN),
Bia\l{}kow (PL), Baja (HU) and ORM (E, Mercator telescope)
observaties.  Preliminary results based on Bia\l{}kow data, enlarged
with older KO data, of h\,Per include the detection of several
variable stars, including three known and six new $\beta$\,Cep stars, some
of them seemingly showing $g$-modes, and eight Be/SPBe stars
(Majewska-\'Swierzbinowicz et al.\,2008).

\subsection*{Discussion}
The open cluster campaign is one of the largest asteroseismic
observational ground-based efforts currently executed, with many sites
and small telescopes involved.  Observations, reduction and analysis are in
progress. So far, the campaign has been very succesfull with several
$\beta$\,Cep, SPB and Be stars already discovered from isolated
datasets.  The combined multi-site, multi-colour time-series, free
from aliasing effects, promise the detection and identification of
several pulsation modes. Moreover, the discovered pulsators are
promising targets for future seismic modeling, given the constraints
on age and chemical composition dictated by the clustership member. 

It seems to be not uncommon to observe high-order $g$-modes in $\beta$
Cep stars. This follows from recent results on the cluster data,
e.g.\, NGC\,6910~38 (Pigulski 2008) and some $\beta$\,Cep variables in
h\,Per (Majewska-\'Swierzbinowicz et al.\,2008). Also other examples are
known (e.g. the CoRoT target HD\,180642, see above). Recently, Miglio et
al. (2007) theoretically predicted high-order modes to be unstable in
$\beta$\,Cep stars, which leaves room for nice prospects in
asteroseismic studies, as the simultaneous detection of $g$- and
$p$-modes enables the mapping of stellar interiors from the core to
the surface.

\section*{Time-scales of extra line-broadening in massive OB stars}
Spectroscopic studies show that a non-negligible extra
line-broadening, often called macro-turbulence, plays an important
role in the line broadening of OB giants and supergiants.  The origin
of the broadening, however, is still unknown (e.g. Sim\'on-D\'{\i}az
\& Herrero 2006). One of the possible explanations is in terms of
stellar oscillations (e.g.\, Aerts 2009, these proceedings). Currently a new
observational project is set up by Sim\'on-D\'{\i}az et al.\, to
investigate this explanation, to explore other origins, to study
whether the macro-turbulence in massive OB stars show variability, and
if so, to investigate on what time-scales it does.
Five nights have been awarded in November 2008 with FIES@Nordic
Optical Telescope in low-resolution mode (R=25000). Eight OB targets
have been selected, and will be observed in different time-intervals,
ranging from a few minutes to 3 days. 
This project is a good step in the direction of exploring possibilities of pulsational and asteroseismic studies of more evolved OB stars.

\acknowledgments{The CoRoT ground-based support observations benefit
greatly from time on ESO Telescopes at the La Silla Observatory under
the ESO Large Programmes LP178.D-0361 and LP182.D-0356 (PI:
E. Poretti).  KU acknowledges Andrzej Pigulski, Sophie Saesen, Juan
Guti\'errez-Soto, Coralie Neiner, Michele Floquet, Conny Aerts, and
Artemio Herrero for their input on the latest results on the projects
described in this paper. She also thanks Ennio Poretti for careful
reading of the manuscript.  
Part of this work has been carried out at INAF-OA
Brera (Merate, Italy) in the framework of a Marie Curie Intra-European
Fellowship, contract number MEIF-CT-2006-024476, and of the Italian
ESS project ASI/INAF/ I/015/07/0, WP 03170.}

\References{
\rfr Aerts C., 2009, CoAst, in press 

\rfr Briquet M., Uytterhoeven K., Aerts C., et al. 2009, CoAst, in press 

\rfr Guti\'errez-Soto J., Fabregat J., Suso J., et al. 2007, A\&A, 476, 927

\rfr Handler G., Tuvikene T., Lorenz D.,  et al. 2007, CoAst, 150, 193

\rfr Majewska-\'Swierzbinowicz A., Pigulski A., Szabo\'o R., \& Csubry Z. 2008, JPhCS, in press

\rfr Miglio A., Montalban J., \& Dupret M.-A.  2007, MNRAS, 375, 21

\rfr Pigulski A., Handler G., Michalska G., et al. 2007, CoAst, 150, 191

\rfr Pigulski A. 2008,  JPhCS, in press

\rfr Poretti E., Garrido R., Amado P.J., et al. 2003, A\&A, 406, 203

\rfr Poretti E., Alonso R., Amado P.J., et al. 2005, AJ, 129, 2461

\rfr Saesen S., Pigulski A., Carrier F., et al. 2008, JPhCS, in press

\rfr Saesen S., Carrier F., Pigulski A., et al.  2009, CoAst, in press

\rfr Sim\'on-D\'{\i}az S. \& Herrero A., 2006, A\&A 468, 1063

\rfr Uytterhoeven K., Poretti E., Rainer M., et al. 2008, JPhCS, in press

\rfr Uytterhoeven K.,  Mathias P., Poretti E., et al. 2008, A\&A, in press (arXiv:0807.0904)

}

\end{document}